%% file: 00_main.tex
\documentclass[preprintnumbers,aps,prd,twocolumn,showpacs,showkeys,floatfix,preprintnumbers,letterpaper,amsmath,amssymb,superscriptaddress]{revtex4-1}

\usepackage{amsmath}
\usepackage{dcolumn}% Align table columns on decimal point
\usepackage{color}
\usepackage{graphicx}
\usepackage{subfigure}
\usepackage{bm}
\usepackage[mathlines]{lineno}
\usepackage{dcolumn}
\usepackage{color}

\def\rhobar{\bar{\rho}}

\begin{document}

\title{Monge-Ampère gravity, optimal transport theory and their link to the Galileons}

\author{Albert Bonnefous}
\email{albert.bonnefous@ens-lyon.fr}
\affiliation{Department of Physics, École Normale Supérieure de Lyon, 69364, Lyon, France}

\author{Yann Brenier}
\email{yann.brenier@ens.fr}
\affiliation{CNRS, École Normale Supérieure, 45 rue d’Ulm, F-75230 Paris cedex 05, France}

\author{Roya Mohayaee}
\email{mohayaee@iap.fr}
\affiliation{Sorbonne Université, CNRS, Institut d'Astrophysique de Paris, 98bis Bld Arago, 75014 Paris, France}
\affiliation{Rudolf Peierls Centre for Theoretical Physics, University of Oxford, Parks Road, Oxford OX1 3PU, United Kingdom}

\begin{abstract}
Mathematicians have been proposing for sometimes that Monge-Ampère equation, a nonlinear generalization of the Poisson equation, where trace of the Hessian is replaced by its determinant, provides an alternative non-relativistic description of gravity. Monge-Ampère equation is affine invariant, has rich geometric properties, connects to optimal transport theory, and remains bounded at short distances. Monge-Ampère gravity, that uses a slightly different form of the Monge-Ampère equation, naturally emerges through the application of large-deviation principle to a Brownian system of indistinguishable and independent particles. In this work we provide a physical formulation of this mathematical model, study its theoretical viability and confront it with observations. We show that Monge-Ampère gravity cannot replace the Newtonian gravity as it does not withstand the solar-system test. We then show that Monge-Ampère gravity can describe a scalar field, often evoked in modified theories of gravity such as Galileons. We show that Monge-Ampère gravity, as a nonlinear model of a new scalar field, is screened at short distances, and behaves differently from Newtonian gravity above galactic scales but approaches it asymptotically. Finally, we write a relativistic Lagrangian for Monge-Ampère gravity in flat space time, which is the field equation of a sum of the Lagrangians of all Galileons. We also show how the Monge-Ampère equation can be obtained from the fully covariant Lagrangian of quartic Galileon in the static limit. The connection between optimal transport theory and modified theories of gravity with second-order field equations, unravelled here, remains a promising domain to further explore.

\end{abstract}

\maketitle

\section{\label{sec:intro}Introduction}
\input{01_Intro}

\section{\label{sec:math} Mathematical background : the Monge-Ampère equation}
\input{02_MA_eq}

\section{\label{sec:MAG_replace} Can Monge-Ampère gravity {\it replace} the Newtonian gravity?}
\input{03_MA_gravity}

\section{\label{sec:MAG_complement} Can Monge-Ampère gravity {\it complement} Newtonian gravity ?}
\input{04_MAG_new_int}

\section{\label{sec:apply_sol} Applying MAG to the Solar system and beyond}
\input{05_Solar_system}

\section{\label{sec:MAG_relat} Relativistic theory of Monge-Ampère equation and MAG}
\input{06_relat_theory}

\section{\label{sec:conclusion} Conclusion}
\input{07_Conclusion}

\section{\label{appendix:perihelion}
Appendix: on the calculation of the advancement of perihelion in Monge-Ampère gravity}
\input{08_Appendix}

\section{Appendix on the optimal transport}
\input{09_Appendix2}

\bibliography{00_main}

\end{document}

%% file: 01_Intro.tex
The classical formulation of gravity based on the Poisson equation has shortcomings and it is well-known that the theory of General Relativity (GR) provides a more complete description of gravity. At the scale of solar system, the theory works so unbeatably well that any other theory has to reduce to GR if it is to be considered a viable theory of gravity \cite{cliff}. However, there is no reason to believe that a single theory would provide a description of dynamics of the Universe at thousands of decades in scale and there are reasons to believe that GR is not the ultimate theory of gravity \citep{clifton_modified_2012,Ishak_2018}. It is not only that we still haven't found a quantum theory of gravity, but also that certain phenomena at cosmological scales urge for a new theory. Indeed, the cosmological constant problem, perhaps the biggest problem in physics \cite{weinberg}, has been a driving motivation for many years behind models of modified gravity \citep{amendola_cosmology_2013}.

Recently, it has been suggested that at the non-relativistic level,  Monge-Ampère equation might provide a better model of gravity than the Poisson equation  \cite{brenier_modified_2011}. Monge-Ampère equation which dates back to two centuries ago has been extensively studied in mathematics (see {\it e.g.}\ \cite{caffarelli,caffarellimilman,figalli,trudinger}) and has recently attracted much attention due to its connection with optimal transport (OT) theory \cite{yann_decompositionpolair2,yann_decompositionpolair,figalli2}. This equation has been used in various branches of science. A notable example is meteorology, where it replaces the Poisson equation in the semi-geostrophic equations that describe successfully the formation of anisotropic structures such as fronts and shocks in the atmosphere \cite{cullenBook}. In astrophysics, Monge-Ampère equation arises as a consequence of mass conservation (3-dimensional) in the Universe. The gradient flow which is an essential linking bridge between the Monge-Ampère equation and the optimal transport theory is a natural consequence of the expansion of the Universe which imposes the rapid decay of  vorticity mode on large scales. Consequently, the Monge-Ampère equation  has provided a mathematically-solid framework to solve the inverse problem of reconstruction in cosmology for the past twenty years (see {\it e.g.} \cite{uriel,brenier_reconstruction_2003,roya1,roya2,brunoroyasebastian,sebastian,farnik1,farnik2,farnik3}).

In addition to the inverse problem of cosmological reconstruction, Yann Brenier has been proposing for some times that the Monge-Ampère equation can also provide a good effective model for the forward dynamics of the Universe and can describe the formation and evolution of large-scale structures in cosmology \cite{brenier_modified_2011,loeper}. In more precise terms, the Monge-Ampère equation might provide a {\it good substitute for the Poisson equation} to describe gravity, at cosmological scales \citep{loeper,brenier_modified_2011,yann2016,ambrosio}. In Brenier's theory, Monge-Ampère gravity (MAG) emerges from a Brownian system of independent, identical and indistinguishable particles through the application of large-deviation principle \cite{brenier_double_2015,ambrosio}. This beautiful mathematical construct has not yet been confronted with observations nor has gone through the basic theoretical symmetry and consistency checks. This is the purpose of the present paper.

MAG is interesting from different aspects. First of all, the Monge-Ampère equation introduces anisotropic tidal tensor through off-diagonal terms,  which could provide a natural modelling of anisotropies in the universe, as it successfully does through semi-geostrophic equations in meteorology where Poisson equation is substituted by the Monge-Ampère equation \citep{cullenBook}. Moreover, Monge-Ampère equation arises in certain Galileon theories (see {\it e.g.}\ \cite{fairlie_comments_2011} and discussions in Section \ref{sec:MAG_relat}). These theories describe a scalar field that adds a new long-range interaction to general relativity, with the fundamental propriety that they are invariant under translation of the form:
\begin{equation}
    \varphi\rightarrow\varphi+a^\mu x_\mu +b\,,
\end{equation}
where $a^\mu$ is any 4-vector, $\varphi$ is a scalar field and $b$ is a constant. In such models with a {\it fifth force}, the full action is that of GR plus the scalar field for which the corresponding field equations are  only second-order in derivatives and of Monge-Ampère type (see {\it e.g.} \cite{fairlie_comments_2011,nicolis_galileon_2009}) .

In this article, we will first consider MAG as a classical and non-relativistic limit of "a model" of gravity, {\it i.e.} as {\it a post-Newtonian} but not-relativistic correction to the classical Newtonian gravity. After laying down some mathematical background in section \ref{sec:math}, we demonstrate in section \ref{sec:MAG_replace} that MAG cannot replace Poisson equation and demonstrate explicitly that it fails the solar system test. Then we consider in section \ref{sec:MAG_complement} MAG as an addition to the classical gravitation, and then test it in the solar system in section \ref{sec:apply_sol}. Finally, we briefly explore the compatibility of this model with a relativistic Lagrangian description and its relation to models of modified gravity such as Galileons in section \ref{sec:MAG_relat}. We review Monge-Ampère equation and its relation to optimal transport theory in the appendix.

%% file: 02_MA_eq.tex
MA equation, in its simplest form, links a potential $\Phi(\mathbf{x})$ to a source density $\rho(\mathbf{x})$ as:
\begin{equation}
    \text{det}(D^2\Phi)=\alpha\rho(\mathbf{x})\,,
    \label{eq:MA_naive}
\end{equation}
where $D^2\Phi = \partial_i\partial_j\Phi$ is the Hessian matrix of $\Phi$, and the constant $\alpha$ is necessary to keep the correct physical units. This equation is highly non-linear and has a large, general linear or affine, symmetry group (\cite{caffarelli}). Hereafter, we refer to this equation as the Monge-Ampère (MA) equation.

Monge-Ampère equation is elliptic degenerate if $\Phi$ is convex. The prototype domain where Monge-Ampère equation arises is what is known as the {\it Minkowski problem}, which is basically the problem of finding a manifold (or a surface) with prescribed boundary and Gaussian curvature \cite{figalli,trudinger}. The Monge-Ampère equation prescribes the product of the eigen-values of the Hessian of $\Phi$, in contrast to the Poisson equation, $\nabla^2\Phi=\rho$ which prescribes the sum of the eigenvalues of $\Phi$.

The Monge-Ampère equation has diverse applications, in mathematics, for instance in affine and convex geometry. Recently, the Monge-Ampère equation has found important applications in optimal transport theory and diverse areas of applied science such as economics and meteorology (see Appendix \ref{appendix:MA-OT},  for historical notes and also the relation between Monge-Ampère equation and optimal transport theory.)

In mathematics and astrophysics, the following version of MA equation:
\begin{equation}
    \text{det}(\mathbb{I}+\beta D^2\varphi_{\rm MA})=\alpha\rho(\mathbf{x})\,,
    \label{eq:MA_varphi}
\end{equation}
which makes a natural correspondence with Poisson equation, is often used, where $\beta$ is a new physical constant. This is the form of the Monge-Ampère equation that has been derived in \cite{brenier_modified_2011}. We refer to this form of the Monge-Ampère equation as the Monge-Ampère-gravity (MAG) equation, or simply the MAG equation. Equations (\ref{eq:MA_naive}) and (\ref{eq:MA_varphi}) are related through the transformation of the potential 
\begin{equation}
\Phi=\frac{\mathbf{x}^2}{2} + \beta\varphi_{\rm MA}\,,  \label{phivarphi}
\end{equation}
with $\mathbf{x}^2/2$ a harmonic potential with an arbitrary origin. The derivation of both of these equations in the case of the OT theory is given in the first part of appendix \ref{appendix:MA-OT} for completeness.

By developing the determinant in equation (\ref{eq:MA_varphi}), we can prove that:
\begin{equation}
    \text{det}(\mathbb{I}+\beta D^2\varphi_{\rm MA})=1+\beta\Delta\varphi_{\rm MA}+\mathcal{O}(\beta^2\varphi_{\rm MA}^2)\,,
    \label{eq:MA_develop}
\end{equation}
which might demonstrate that, the Poisson equation can be formally considered as a weak field limit of the MA equation. On the contrary, when $\beta\,D^2\varphi_{\rm MA}\gg 1$, for the same harmonic potential, then equation (\ref{eq:MA_varphi}) approach equation (\ref{eq:MA_naive}). Therefore the MAG equation (\ref{eq:MA_varphi}) can be seen as some sort of intermediate between the Poisson equation and the {\it usual} MA equation (\ref{eq:MA_naive}).

We can show fairly easily that equation (\ref{eq:MA_varphi}) is translationally and rotationally invariant. However, this equation is no longer linear in $\varphi_{\rm MA}$ and scale-invariant contrary to the Poisson equation. Still, a characteristic scale  appear through the combination of the mass of the source and the background density (as we shall show in expression \eqref{eq:R}) and doesn't appear in the usual manner, as for example it does in the Yukawa coupling potential, since in this form $\alpha$ has the dimension of the inverse of a density, and $\beta$ has the dimension of a squared time. The original MA equation (\ref{eq:MA_naive}) is also unimodular affine invariant \citep{caffarelli}. However it is not formally true for the potential $\varphi_{\rm MA}$ in equation (\ref{eq:MA_varphi}), where the same affine-invariance is still present but modulo the harmonic potential.

A simple dimensional analysis implies that $\alpha$ has the dimension of an inverse mass density and we choose to write it as $\alpha={1/\rho_{\rm MA}}$. Again a dimensional analysis suggests that $\beta$ has the dimension of time-squared, which could also be written as $(4\pi G_{\rm MA}\rho_{\rm MA})^{-1}$, with $G_{\rm MA}$ a new constant with the same units as Newton's gravitational constant, $\text{m}^3.\text{kg}^{-1}.\text{s}^{-2}$. This parameter is similar to the inverse of the Hubble constant squared. With this new choice of parameters $\rho_{\rm MA}$ and $G_{\rm MA}$, we can express the MAG equation as:
\begin{equation}
    \text{det}(\mathbb{I}+(4\pi G_{\rm MA}\rho_{\rm MA})^{-1}D^2\varphi_{\rm MA}) = \rho/\rho_{\rm MA}\,.
    \label{eq:MAG}
\end{equation}
Note that these two parameters can be functions of time since we are not preoccupied with dynamics. We chose not to absorb the constants in the field $\varphi_{\rm MA}$ in order to keep a coherence with the usual formulation of Poisson equation for gravity. It is a usual practice to absorb $4\pi\rho_{\rm MA}$ in $G_{\rm MA}$, but for clarity we leave this parameter explicit. With these results we can already make some interpretations of these two parameters. With equation (\ref{eq:MAG}), we can see that the approximation (\ref{eq:MA_develop}) can be made if $\rho/\rho_{\rm MA}\approx1+o(1)$. Therefore, $\rho_{\rm MA}$ is the characteristic density where this approximation can be made when $\rho-\rho_{\rm MA}\ll \rho_{\rm MA}$. When this parameter is fixed, $\varphi$ is proportional to $G_{\rm MA}$. Therefore this parameters represents the characteristic strength of the field.

%% file: 03_MA_gravity.tex
Hereafter, we shall write $\varphi_{\rm N}$ for the classical Newtonian potential that satisfies the Poisson equation, $\varphi_{\rm MA}$ is still the potential that satisfies the MAG equation (\ref{eq:MAG}). We recall that our aim is to see if Monge-Ampère gravity can provide a non-relativistic, post-Newtonian-type, replacement to classical Newtonian gravity as is implied by \cite{brenier_modified_2011}. To serve this purpose, the MA equation would need to be coherent with Poisson equation for the gravitational field:
\begin{equation}
    \Delta\varphi_{\rm N} = 4\pi G_{\rm N}\rho \,, 
\end{equation}
where $G_{\rm N}$ is the gravitational constant. In cosmology a slightly different form of Poisson equation is used, introducing the mean density of the universe $\rhobar$:
\begin{equation}
    \Delta\varphi_{\rm N} = 4\pi G_{\rm N}(\rho-\bar{\rho})\,.
    \label{eq:poisson}
\end{equation}
Since $\rho\gg\rhobar = 9.9\cdot  10^{-30}\ \text{g.cm}^{-3}$ for most physical systems at galaxy scales, this background term can be ignored most of the times. In a viable physical theory, a global density in a local equation may seem structurally wrong. However, the introduction of $\rhobar$ is formally necessary to ensure that no gravitational potential is created from an uniform distribution. It can also be justified by considering the Poisson equation in comoving coordinates in an expanding Friedmann Universe \cite{peebles}. Although, we comment that this form of the Poisson equation is only a mathematical approximation to allow a simple description of gravity at cosmological scale, and is inappropriate for a fundamental description of any interaction. 

In \cite{brenier_modified_2011}, it is implied that $\rho_{\rm MA}=\rhobar$, and we observe that in the approximation (\ref{eq:MA_develop}), we can recover the Poisson equation in cosmology (\ref{eq:poisson}) with $G_{\rm MA}=G_{\rm N}$. If we then require that the Newtonian gravity is recovered at short distances, {\it e.g.} for the Earth, with a mean density of $\rho_{\rm earth}=5.5\ \text{g.cm}^{-3}$, we have $\rho_{\rm earth} \gg\bar{\rho}$, the right side of the equation (\ref{eq:MAG}) is far greater than 1, and the approximation (\ref{eq:MA_develop}) cannot be made.

The only solution would be to replace $\bar{\rho}$ by a new characteristic density $\rho_{\rm MA}$. But to keep the Newtonian approximation inside the solar system, this $\rho_{\rm MA}$ which would be {\it huge} compared to the characteristic density of the Sun and the different planets.
Furthermore, our calculations of the perihelion advancements of different solar system planets show that, each time a different value of $\rho_{\rm MA}$ is needed to fit the observations. We shall not detail these calculations here because the comparison of this large background density to the density of the sun and the planets and the scaling argument presented earlier in this subsection already rule out this scenario.

We can thus conclude that MAG cannot replace the Poisson gravity as it will not pass the solar system test. A very large "varying" density is required which would render the theory unfeasible. Furthermore, a large unphysical background density  would imply that MAG will not provide a good theory of large-scale structure formation at cosmological scales and hence it would lose its relevance to optimal transport theory, which requires that $\rho=\bar\rho$, and which is an appealing aspect of Monge-Ampère equation.

In the next section we consider a physically-motivated case, one in which the Poisson equation is not {\it replaced} but rather {\it complemented} by the Monge-Ampère equation, which describes a scalar field and arises naturally in different models of modified gravity \cite{clifton_modified_2012}. 

%% file: 04_MAG_new_int.tex
In the previous section, we have shown that MA cannot replace Poisson equation in describing gravity in the weak field limit. In this section, we explore the idea that MA equation can describe a new scalar field which complements the gravitational effect of the matter field described by GR. 

If we consider MA equation as an equation of a scalar field then there is no necessity to link it to the Poisson equation, and hence the parameter $G_{\rm MA}$ does not need to be identical to the gravitational constant $G_{\rm N}$. However, to stay consistent with OT theory, we choose $\rho_{\rm MA}=\rhobar$. The mechanism derived in \cite{brenier_modified_2011} would describe an emergent interaction that would be additional to the effect of gravity. We refer the reader to \cite{ambrosio,MAGus} for detailed derivation of MAG from statistical mechanics, namely from a Brownian system of independent and indistinguishable particles.

In the non-relativistic limit, the dynamics of matter, of $\varphi_N$ and $\varphi_{\rm MA}$ are then given by:
\begin{equation}
    \Delta\varphi_{\rm N} = 4\pi G_{\rm N}\delta\rho\,,
    \label{eq:Poisson_cosmo}
\end{equation}
\begin{equation}
    \text{det}(\mathbb{I}+(4\pi G_{\rm MA}\rhobar)^{-1}D^2\varphi_{\rm MA}) = (\rhobar+\delta\rho)/\rhobar\,,
\end{equation}
\begin{equation}
    \frac{\text{d}\mathbf{v}}{\text{d}t} = -\nabla\varphi_{\rm N}-\nabla\varphi_{\rm MA}\,,
    \label{eq:NII}
\end{equation}
\begin{equation}
    \frac{\text{d}\rho}{\text{d}t}+\nabla\cdot\mathbf{v}=0\,,
\end{equation}
with $\rho=\rhobar+\delta\rho$. Now, we study this system of equations for two simple cases of static mass density, the point mass and the sphere of constant density. These two density profiles are good models for most spherical sources such as planets, stars or galaxies in certain problems.

\subsection{Resolving MAG for spherically-symmetric densities}
The non-linearity of the elliptic Monge-Ampère equation makes it extremely difficult to solve. However its direct relation to optimal transport theory has lead to the development of powerful algorithms, namely discrete-discrete entropic regularisation and semi-discrete optimal transport which are used to solve this nonlinear equation numerically \citep{merigot,cuturi,journals/M2AN/LevyNAL15,peyre-cuturi,levy_mnras_2021,bertsekas}. 

However, for spherically symmetric systems, analytic solutions can be found. In this article we shall deal with spherically symmetric densities $\rho(r)$. The spherically symmetric form of the Monge-Ampère equation:
\begin{equation}
    \frac{1}{r^2}(\partial_r\Phi )^2\partial_r^2\Phi=\frac{1}{3r^2}\partial_r(\partial_r\Phi)^3=\alpha\rho(r)\,,
    \label{eq:MA_spherical}
\end{equation}
can be integrated for any spherically symmetric source. However, since we can only have an integral expression for $\partial_r\Phi$ with equation (\ref{eq:MA_spherical}), we shall integrate from $0$ to $r$, provided that the density profile is regular enough at $r=0$. Note that this is not the same as the usual practice for the Poisson equation which is integrated from $r$ to $\infty$, with the boundary condition that $\partial_r\varphi_{\rm N}\rightarrow0$ at $r\rightarrow\infty
$. With $\beta\varphi_{\rm MA}=\Phi-r^2/2$ we have:
\begin{equation}
    \beta\partial_r\varphi_{\rm MA}=\partial_r\Phi(r) -r=\left(\int_0^{r}  3\alpha\rho(r') r'^2\text{d}r' +C^3\right)^{1/3}-r\,,
    \label{eq:MA_spherical_resolved}
\end{equation}
with $C=\partial_r\Phi(r=0)=\beta\partial_r\varphi_{\rm MA}(r=0)$ an integration constant. To determine the value of this we can use a symmetry argument. The potential $\varphi_{\rm MA}$ should keep the spherical symmetry of the density profile, and provided that there is no irregularities at $r=0$ we should have $\partial_r\varphi_{\rm MA}(r=0)=0$. Therefore, we obtain that $C=0$. Now, using equation (\ref{eq:MA_spherical_resolved}), with the correct values for $\alpha$ and $\beta$, and a density $\rho=\rhobar+\delta\rho$, we obtain the following expression for $\varphi_{\rm MA}$:
\begin{equation}
    \partial_r\varphi_{\rm MA}=4\pi G_{\rm MA}\rhobar\left(\left(r^3+\int_0^{r}  3\frac{\delta\rho(r')}{\rhobar}r'^2\text{d}r' \right)^{1/3}-r\right)\,.
\label{eq:MA_spherical_resolved2}
\end{equation}
The value of the potential $\varphi_{\rm MA}$ can then be obtained by direct integration of this result. However since for our study only the gradient of the potential is important we shall not solve the above expression to find specific form of the potential.

\subsection{Solution for a sphere of constant density}
The density of a sphere of radius $r_0$ and constant density $\rho_0=\frac{m}{4/3\pi r_0^3}$ can be written as $\delta\rho(\mathbf{x})=\rho_0H(r/r_0)$, with the notation $H(x) = \{1 \text{ if $x<1$}, 0 \text{ if $x\geq 1$}\}$. Using equation (\ref{eq:MA_spherical_resolved2}) for this density we find that:
\begin{equation}
    \partial_r\varphi_{\rm MA} = 
    \begin{cases}
        4\pi G_{\rm MA}\rhobar\, r\left(\left(1+\rho_0/\rhobar\right)^{1/3}-1\right) & \text{if $r<r_0$} \\
        4\pi G_{\rm MA}\rhobar\, \left(\left(r^3+\frac{3m}{4\pi\rhobar} \right)^{1/3}-r\right) & \text{if $r>r_0$}
    \end{cases} 
\end{equation}
This sphere of constant density is a simple model for any spherical object than can be improved by considering a non-constant density profile along its radius $\rho(r)=m\frac{\sigma(r)}{r^2}$, such that $\int_0^{r_0}\sigma(r)\text{d}r=1$, and $\sigma(r>r_0)=0$. However, by integrating equation (\ref{eq:MA_spherical_resolved}) we find immediately that the potential outside of the sphere doesn't depends on the density profile $\sigma(r)$. Everything act as if there was a point mass $m$ at $r=0$. This is a similar results as the Gauss' law applied to the gravitational field in spherical symmetry, without the linearity of the field equation and the divergence necessary to apply Green-Ostrogradski theorem. Therefore, to understand planetary dynamics we can limit our study to a point-mass source.

\subsection{Solution for a point mass}
A point mass can be seen as a limiting case for a sphere where its radius $r_0$ tends to 0, or equivalently as $r\rightarrow\infty$, while its mass $m$ stays constant. Therefore we have the same expression as the sphere for $r>r_0$, but here for any $r>0$.
\begin{equation}
    \partial_r\varphi_{\rm MA}=4\pi G_{\rm MA}\rhobar\left(\left(r^3+\frac{3m}{4\pi\rhobar}\right)^{1/3}-r\right)\,.
\end{equation}
Note that we evacuated here the continuity issues at the origin with this limiting argument. Indeed we have not rigorously proved that the equation (\ref{eq:MA_spherical_resolved2}) is true for a density $\delta\rho=m\delta(\mathbf{x})$ since this equation is only true for a density profile that is {\it regular}. However since the point mass is only an approximation and not a real physical density, it is unimportant for the purposes of this study. For a more rigorous insight into the resolution of Monge-Ampère equation in spherical symmetry, see \cite{Frisch_2010}. 

Next, we identify $R$ the characteristic scale,
\begin{equation}
    R=\left(\frac{3m}{4\pi\rhobar}\right)^{1/3}\,,
    \label{eq:R}
\end{equation}
of the model, which is consistent with our findings in the previous section that the Monge-Ampère equation, in its form (\ref{eq:MA_varphi}) is no longer scale-invariant. This scale can be interpreted as the radius of the sphere of constant density $\rhobar$ that would have the same mass $m$ as the point mass. The solution becomes:
\begin{equation}
    \partial_r\varphi_{\rm MA}=4\pi G_{\rm MA}\rhobar R\left(\left(1+\frac{r^3}{R^3}\right)^{1/3}-\frac{r}{R}\right)\,.
    \label{eq:MA_point}
\end{equation}
In Fig. \ref{fig:MAGvsN}, the behavior of MAG is compared to its Newtonian counterpart in the case of the gravitational field created by the Sun. At small radii MAG affinely reaches a constant value whereas the Poisson equation diverges; the MA force is screened at short distances. The figure also shows clearly that the Newtonian gravity is recovered at large distances.

\subsubsection{Comparison with Newtonian gravity}
We write the Newtonian potential for a point mass :
\begin{equation}
    \partial_r\varphi_{\rm N} = \frac{G_{\rm N}m}{r^2}\,,
\end{equation}
for comparison with that of MAG. We can already make a few observations. Far from the source, at $r\gg R$, we may expand equation (\ref{eq:MA_point}). The first two dominant terms are:
\begin{equation}
    \partial_r\varphi_{\rm MA} =\frac{G_{\rm MA} m}{r^2} - \frac{G_{\rm MA} m}{3} \frac{R^3}{r^5}+ ...
    \label{eq:MA_perturb}
\end{equation}
The first term is equivalent to the classical Newtonian potential for a point mass. This is coherent with the results that Poisson equation is a weak field limit of the Monge-Ampère equation. However the second term is negative, and therefore MAG is weaker than its Newtonian counterpart, modulo a different constant $G_{\rm MA}$.  Monge-Ampère potential also no longer diverges at $r=0$ like $\partial_r\varphi_{\rm N}$, and its radial gradient becomes constant when $r\ll R$, with a value at $r=0$ of:
\begin{equation}
\partial_r\varphi_{\rm MA}(r=0)=4\pi G_{\rm MA}\rho_{\rm MA}R\,.
\end{equation}
The effect of MAG is screened at short distances. This effect is similar to the Vainshtein screening mechanism of the Galileon \cite{chow_galileon_2009} which arises due to the effect of non-linearities, becoming prominent in strong gravitational field. The screening mechanism renders this {\it the fifth force}-type interaction negligible at high-density regions. We mention that there is an interesting observational test of this mechanism, due to the break down of strong equivalence principle \cite{hui}.

%% file: 05_Solar_system.tex
\begin{figure}
    \centering
    \includegraphics[scale=0.8]{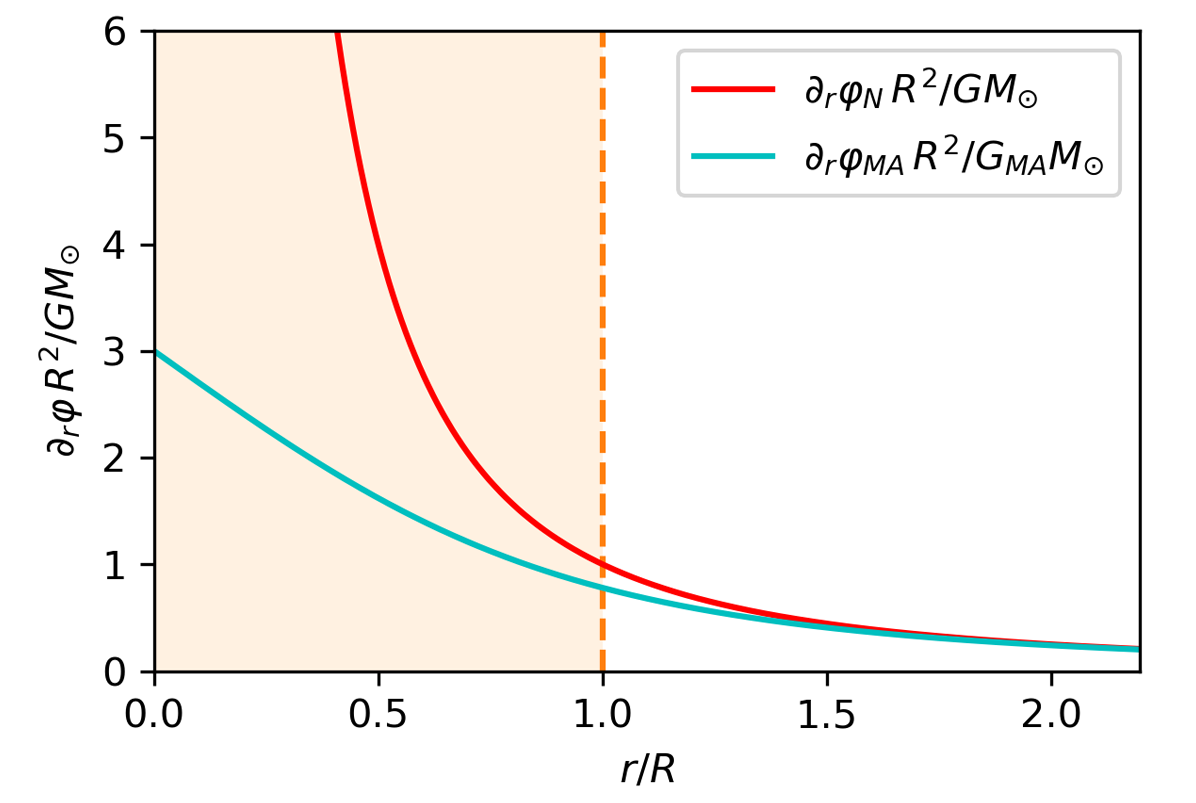}
    \caption{The gravitational force of the Sun is shown, in red for the classical Newtonian gravity and in cyan for Monge-Ampère gravity (MAG). Both of the field gradients have been normalized by $R^2/G_{\rm N}M_\odot$ and $R^2/G_{\rm MA}M_\odot$ respectively. We note that as $\varphi_{MA}$ is proportional to $G_{MA}$, the precise value of the latter doesn't play a role in the normalization. The MAG characteristic scale for the sun is $R=(3M_\odot/4\pi\rhobar)^{1/3}=3.6\cdot10^{18}\text{m}\approx110\text{pc}$. For $r>R$, in white region, we can see that MAG follows closely the behavior of the Newtonian gravity. For $r<R$, in orange region, the behavior of MAG diverges from its classical counterpart and exhibits an affine behavior, $\partial_r\varphi_{\rm MA}\approx4\pi G_{\rm MA}\rhobar(R-r)$. In MAG, very close to the sun, $r\ll R$, the gravitational pull becomes constant $\partial_r\varphi_{\rm MA}\approx4\pi G_{\rm MA}\rhobar R$.
    }
    \label{fig:MAGvsN}
\end{figure}

\subsection{MAG inside the solar system}
For the Sun, with a mass $M_\odot=1.989\cdot10^{30}\ \text{kg}$, we obtain a corresponding MAG characteristic scale (\ref{eq:R}) of $R=(3M_\odot/4\pi\rhobar)^{1/3}=3.6\cdot10^{18}\text{m}$. This scale is far superior to the size of the solar system (the aphelion of Pluto is $7.4\cdot10^{12}\ \text{m}$). We can thus consider that the effects of MAG are fully screened and its contribution is a constant force. Considering both the effect of the Sun's Newtonian gravity and of MAG, the second law of Newton (\ref{eq:NII}) for any point object at the position $\mathbf{r}$ inside the solar system would then become:
\begin{equation}
    \frac{\text{d}^2\mathbf{r}}{\text{d}t^2} = \left(-\frac{G_{\rm N}M_\odot}{r^2}-4\pi G_{\rm MA}\rhobar R\right)\hat{\mathbf{r}}\,.
    \label{eq:MA_solarsystem}
\end{equation}
For this contribution to be compatible with the observations of the dynamics of the Solar system, it has to be small enough to not have been detected yet. Looking at the dynamics of the solar system might give us another higher bound for the value of $G_{\rm MA}$. Since the effect of the Newtonian gravitation are decreasing with $r$ while MAG stays constant, its relative importance becomes higher and it may be expected that this contribution is more visible at high distance.

\subsection{Modification of the perihelion with MA gravity}
Using classical gravitation theory, the solutions to a two-body problem are the conic trajectories. Among these solutions, the only closed trajectories are ellipses and circles. In a first approximation the trajectories of the planets in the solar system can almost be described by closed ellipses. However, when we take into account the gravitational effect of the other planets and the correction of GR, their trajectories are no longer closed. For example, we observe an advance of the perihelion of Mercury, Venus, the Earth, {\it etc}. Since MAG also brings a correction to the classical gravitation theory, it should also brings a modification to the movement of perihelion over time.

Using perturbation theory with equation (\ref{eq:MA_solarsystem}), we find that for each revolution the perihelion of a planet in orbit around the Sun is shifted by an angle of $\delta_{\rm MA}$ which can be expressed in radian as:
\begin{equation}\label{eq:precession_MA}
\delta_{MA}=-8\pi^2\left(\frac{3}{4\pi}\frac{\rhobar^2}{M^2}\right)^{1/3}\frac{G_{\rm MA}}{G_{\rm N}}\frac{(1+e)^{1/2}}{(1-e)^{3/2}}\, r_-^2\,, 
\end{equation}
where $e$ is the eccentricity of the orbit and $r_-$ is the perihelion radius. Details of the calculation are presented in the second part of the appendix \ref{appendix:MA-OT}. As we have explained in the previous subsection, in MAG the modification of the perihelion gets stronger at larger $r_-$, and when the body is farther away from the Sun. Notably, according to MAG, there should be a {\it retreat} of the perihelion. The same phenomena happen within Galileon theories \citep{trodden}. The retreat of the perihelion in MAG is stronger at large distances, which is the opposite of the advance of the perihelion in GR that is stronger when we are close to the Sun. Therefore, we can estimate the highest value of $G_{\rm MA}$ that is compatible with the observation of the solar system with the perihelion of Saturn, the farthest planet of the solar system for which we have reliable information on the perihelion precession. The observed perihelion precession of Saturn, is $0.014\pm 0.002\ ''.\text{cy}^{-1}$, and its theoretical value, taking into account GR and the effect of every other planet is $0.0138\ ''.\text{cy}^{-1}$ \cite{nyambuya_azimuthally_2010}. Since these two values are coherent, for the effect of Monge-Ampère to be compatible with these, the precession (\ref{eq:precession_MA}) needs to be smaller than the uncertainty of $0.002 ''.\text{cy}^{-1}$. This gives us a higher bound of $G_{{\rm MA} max}=7\cdot10^{-8}\ \text{m}^3.\text{kg}^{-1}.\text{s}^{-2}$. This upper limit is larger than the Newton constant $G_{\rm N}$ by an order of $10^3$ and hence should not be a cause of concern for the model.

\subsection{MAG outside the solar system}
For an object that is far away from the solar system, in the Newtonian limit of MAG $r\gg R$, the total effect of the Sun's Newtonian gravity and of MAG become:
\begin{equation}
    \nabla(\varphi_{\rm N}+\varphi_{\rm MA})=\hspace{-0.1cm} \left(\hspace{-0.1cm}(G_{\rm N}+G_{\rm MA})-\frac{G_{\rm MA}}{3}\frac{R^3}{r^3}+\cdots\hspace{-0.1cm}\right)\frac{M_\odot}{r^2}\hat{\mathbf{r}}\,.
\end{equation}
At first order in $r$, the effect of MAG on such (galactic) scales might be a slight modification of the gravitational constant to include the MAG contribution, $G_{\rm N}$ becomes $G_{\rm N}+G_{\rm MA}$. If the effect of MAG at the scale of the solar system is screened, gravity might appear stronger on larger scale. This might for example, help explain a part of the rotation curves of the galaxies.
However, since the MAG equation is highly non-linear, this conclusion might not hold. Further studies are necessary to confirm this effect in the case of anisotropic systems, which are better models for galaxies and shall be pursued in future works.

If this additional contribution to the gravitational constant at large scale is theoretically confirmed, and since no observation yet are leading toward a gravitational constant that would be {\it very} different between the solar system scale, the galactic scale and the cosmological scales we might expect that $G_{\rm MA}<G_{\rm N}$ or $G_{\rm MA}\approx G_{\rm N}$ {\it at least} to be coherent with this observation.

We can notice that this requirement is far smaller than the higher bound for $G_{\rm MA}$ found using the perihelion precession of Saturn. This indicate that the study of MAG in the galactic and cosmological scale could be far more pertinent than the study at the solar system scale.

\subsection{MAG beyond galactic scales}
We can also calculate the value of the scale parameter $R=(3M)/(4\pi\rhobar)^{1/3}$ (\ref{eq:R}) for our own galaxy. Taking the total mass of our galaxy, which is around $10^{12}M_\odot$, we obtain the value of about 1.2 Mpc for $R$ which is about the size of our local group. Below this, the acceleration due to the scalar field is constant and, if we choose $G_{\rm N}=G_{\rm MA}$, has a value of about $3\cdot10^{-13}\text{m}.\text{s}^{-2}$. This is much lower than the upper-bound on a constant force obtained for example by the pioneer spacecraft in the solar system \cite{pioneer0}. We note that this upper-bound was explained by the systematic from the spacecraft itself \cite{pioneer} but however provides a present limit on the magnitude of a constant acceleration in the solar system. Hence the effect of this screening force in the local group is negligible and it is only at scales larger than the local group that the force plays a role. 

\begin{figure}    \includegraphics[width=0.97\columnwidth]
    {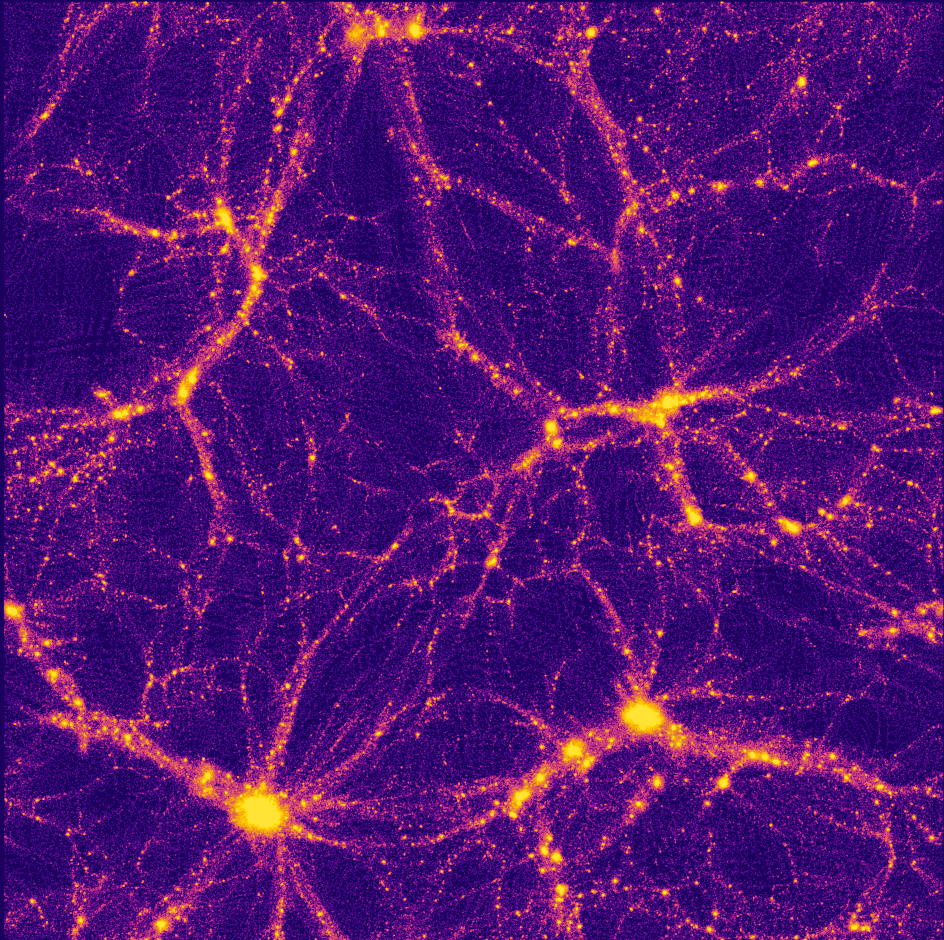}
     \includegraphics[width=0.97\columnwidth]
   {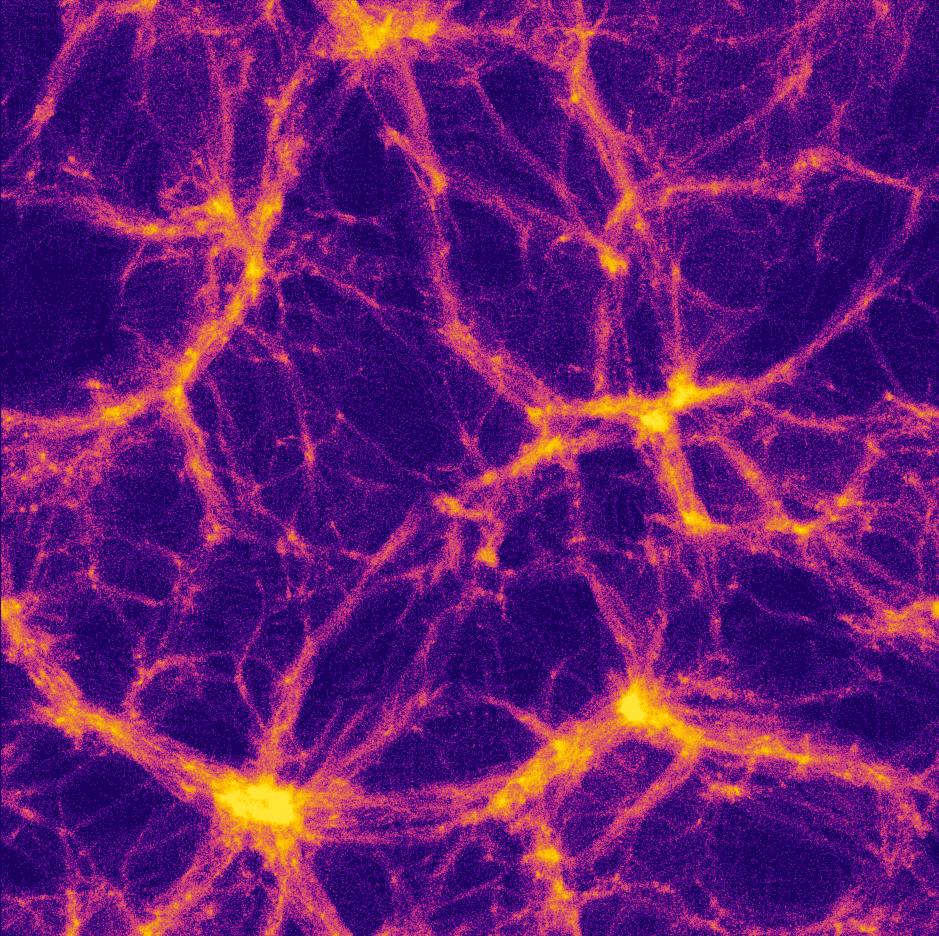}
    \caption{Thin slices  of $256^3$ simulations in boxes of 60 Mpc/h with Newtonian dynamics (top panel) and MAG (bottom panel) with $\Lambda$CDM initial conditions \cite{planck} (See the companion paper \citep{MAGus} where numerical simulations of MAG is presented). The clustering is less prominent in Monge-Ampère because of screening at short distances. This is expected; GR as a spin-2 model of gravity is attractive, and any modification of GR, {\it e.g.}\ with a scalar degree of freedom, would typically yield a repulsive force.}
    \label{fig:mag-newton}
\end{figure}
The screening of MAG in strong field, has also been studied in a companion paper \cite{MAGus} where we have studied MAG as an effective theory at cosmological scales, using novel N-body simulations based on optimal transport theory. Fig. \ref{fig:mag-newton} demonstrates the suppression of clustering in MAG at small scales again due to this screening effect (see \cite{MAGus} for a full account). Indeed, modifications to GR typically yield a repulsive force, which explains the reason for the large interest in such theories as they provide an alternative, to the cosmological constant of GR, explanation of the accelerated expansion of the Universe. We wish to make a comment at this point. Our analytic results and the snapshots in Fig. \ref{fig:mag-newton}, clearly show that there are fewer small halos in MAG than in $\Lambda$CDM Universe. In this work, we have explained this feature in terms of a screening force and have shown that MAG is related to alternative models of gravity, {\it e.g.} Galileons, which naturally encapsulate similar screening mechanisms. However, many different mechanisms for suppression of small-scale structures have been proposed in the literature, among which is the Bose-Einstein or fuzzy dark matter model \cite{tremaine} is which dark matter behaves as a super-fluid and obeys the Schrodinger equation. The decomposition of the wave function yields the continuity and Euler (or Hamilton-Jacobi for gradient velocities) equations, the latter with a quantum pressure term. The quantum pressure term suppresses small scale clustering. It is well-known that {\it Schrodinger problem} is directly related to optimal transport and hence connected to MAG equation \eqref{eq:MA_varphi} \cite{christian}. We shall explore this interesting aspect of MAG and its link to alternative models of dark matter in the forthcoming works.

%% file: 06_relat_theory.tex
In the previous sections, we have considered MAG as a model of gravity in the non-relativistic limit. However, any viable non-relativistic theory of gravity is expected to emerge from a relativistic theory. To enforce the hypothesis that MAG is compatible with the description of a new interaction or emergent mechanism, we need find a Lorentz-invariant Lagrangian that would give a covariant field equation through the Euler-Lagrange variational principle, which in turns would reduce to MAG equation in the non-relativistic limit.

In this section, we first write a relativistic Lagrangian whose field equation in the static limit is the Monge-Ampère equation \eqref{eq:MA_naive}. Second, we write a Lagrangian whose field equation in the static limit is the MAG equation \eqref{eq:MA_varphi}.

\subsection{Monge-Ampère equation as the field equation of quartic Galileon}
There are a large class of Lagrangians whose field equations are second order because such field equations are ghost-free (see {\it e.g.}\ \cite{fairlie1,fairlie2,fairlie_comments_2011,gillescedricdeser,gilles2,fairlie3,deserfradkin}). Writing a Lagrangian that corresponds to such a equation of motion which only depends on the second derivatives of the field is not complicated. Using the elegant formulation given by \cite{fairlie1,fairlie2,fairlie_comments_2011}, one can write in a compact form the equations of motion as:
\begin{equation}
    \mathcal{E}(\partial\partial\varphi)=\rho\,,
\end{equation}
with $\mathcal{E}$ here is a determinant acting only on the second derivatives of the field $\varphi$. This field equation can be obtained through the Euler-Lagrange variational principle from the Lagrangian:
\begin{equation}\label{eq:lagrangian_E}
    \mathcal{L}=\mathcal{E}(\partial\partial\varphi)\varphi-\rho\varphi\,.
\end{equation}
which is valid for the special case of $\mathcal{E}$ being a determinant and apparently many other general forms of this operator \cite{fairlie_comments_2011}.

The MA equation (\ref{eq:MA_naive}) is obtained when the operator $\mathcal{E}$ is a 3-dimensional determinant, and hence it has the following direct Lorentz-invariant equivalent in 4 dimensions:
\begin{equation}
    \text{det}(\partial_\mu\partial_\nu\varphi)=\rho\,.
    \label{eq:MA4d}
\end{equation}
This equation is the field equation of the quintic Galileon ${\cal L}_5$. Note that the Lagrangian obtained using equation (\ref{eq:lagrangian_E}) with this field equation is not  the Lagrangian that is sometimes written for the quintic Galileon. An elegant formulation for the Lagrangian density for quintic Galileon is \cite{gillescedricdeser}:
\begin{equation}\label{eq:lagrangian_5}
    {\cal L}_5=-
\epsilon^{\mu\nu\rho\sigma}\epsilon^{\alpha\beta\gamma\delta} \varphi_{,\mu}\varphi_{,\alpha}\varphi_{,\nu\beta}\varphi_{,\rho\gamma}\varphi_{,\sigma\delta}-\rho\varphi\,,
\end{equation}
where we have also added the last term which is the coupling to the matter field. In the above expression, $\epsilon$ are totally anti-symmetric Levi-Civita tensors and indices of the scalar field denote derivatives, {\it e.g.}\ $\varphi_{,\mu\nu}=\partial_\mu\partial_\nu\varphi$. Through Euler-Lagrange variational principle the above Lagrangian density results in the same equation of motion  as \eqref{eq:MA4d} (see also \cite{fairlie_comments_2011}). 

In fact, Galileons are a subclass of Horndeski theories \cite{horndeski} (and see \cite{deffayetsteer,horndeskireview} for reviews), which describe all scalar-tensor theories whose field equations are second order in curved 4-dimensional space-time. Indeed, it is remarkable that the general action for a "maximal" Galileon in $D$ dimensions, that is to say the Galileon of order $D+1$, can be written as \cite{gillescedricdeser,gilles2}:
\begin{eqnarray}
I&=&\int {\rm d}^D x \,{\cal L}_{D+1}\nonumber\\
&=&\int {\rm d}^D x \varphi\left[\epsilon^{\mu\nu\cdots\alpha}\epsilon^{\gamma\delta\cdots\beta}
\varphi_{,\mu\gamma}\,\varphi_{,\nu\delta}\cdots\varphi_{,\alpha\beta}
\right] -\rho\,\varphi \,\nonumber\\
&=&\int \, {\rm d}^Dx\,\varphi \,{\rm det}(\varphi_{,\alpha\beta}) -\rho\,\varphi \,\,.
\end{eqnarray}
This means that, in  3-dimensions, the quintic Galileon becomes null and we cannot obtain 3-dimensional MA equation \eqref{eq:MA_naive} as a static limit of quintic Galileon.

However, the field equation of the quartic Galileon, ${\cal L}_4$,  can reduces to the 3-dimensional MA equation. The equation of the quartic Galileon involves the operator:
\begin{equation}
    \mathcal{E}_4(\partial\partial\varphi)=\sum\text{det}_3(\partial_\mu\partial_\nu\varphi)\,,
\end{equation}
where $\sum\text{det}_3$ is the linear combination of all the 3 by 3 Hessian matrices for all the possible subset of the variables $\{t,x,y,z\}$ \cite{fairlie_comments_2011}. This equation has the advantage that it doesn't become null in the static case, and instead reduces to the original 3-dimensional MA equation. The field equation of cubic Galileon, ${\cal L}_3$, involves a sum of determinant of 2 by 2 matrices, and cannot be reduced to a Monge-Ampère equation in 3-dimensions.

This sort of hierarchy repeats itself in 2 dimensions, where in the static limit the quartic Galileon becomes null and the field equation of the cubic Galileon, ${\cal L}_3$, reduces to a 2-dimensional MA-{\it like} equation. 

We have shown that quartic Galileon directly yields Monge-Ampère equation \eqref{eq:MA_naive} in static limit and holds in all backgrounds. We note that the quartic Galileon together with ${\cal L}_2$ Galileon, have rather interesting properties; they admit a Schwarzchild-de Sitter exact solutions, pass the solar-system test and yield a small value for the cosmological constant (see \cite{gillbabichev} for further details). 

%------
\subsection{Monge-Ampère gravity (MAG) as the field equation of the sum of all Galileons in Minkowski space}
%-------

Next we write a relativistic Lagrangian, in flat space, corresponding to the Monge-Ampère gravity equation used in MAG \eqref{eq:MA_varphi}. We can formally replace the Laplacian of the identity map plus the Hessian in \eqref{eq:MA_varphi} by
\begin{equation}\label{eq:transfo_psi_phi}
    \partial_\mu\partial_\nu\varphi\rightarrow\eta_{\mu\nu}+\partial_\mu\partial_\nu\varphi\,,
\end{equation}
while keeping the other derivatives the same, which corresponds to relativistic MAG field equation, with a signature $\eta_{\mu\nu}=(-,+,+,+)$ :
\begin{equation}\label{eq:MAG_relat}
    -\text{det}(\eta_{\mu\nu}+\partial_\mu\partial_\nu\varphi)\,=\rho\,.
\end{equation}
In the static case ($\partial_t=0$), the MA operator becomes, once again for $\eta_{\mu\nu}=(-,+,+,+)$:
\begin{equation}
\begin{split}
    -\text{det}(\eta_{\mu\nu}+\partial_\mu\partial_\nu\varphi) & =-\eta_{00}\text{det}(\mathbb{I}+D^2\varphi)\\
    & = +\text{det}(\mathbb{I}+D^2\varphi)\\
\end{split}\,,
\end{equation}
which is the Monge-Ampère gravity equation in 3 dimensions that we had started with \eqref{eq:MA_varphi}. The relativistic field $\varphi$ then corresponds to the classical field $\varphi_{\rm MA}$ of MAG. We note that the equation (\ref{eq:MAG_relat}) assumes that when we change the convention for the Minkowski metric, {\it i.e.} $\eta_{\mu\nu}\rightarrow -\eta_{\mu\nu}$ then we can also redefine $\varphi\rightarrow -\varphi$ and hence the determinant is invariant under the sign convention.

Next, we seek a Lagrangian for MAG in terms of the Galileons. Next, we show that a particular linear combination of all the Galileons yield a Lagrangian forming this field equation \eqref{eq:MAG_relat} through Euler-Lagrange transformation. This Lagrangian is, using the notations of \cite{gillescedricdeser}:
\begin{equation}\label{eq:L_MAG}
    {\mathcal L}_{\rm MAG}=-\sum_{n=1}^5\frac{{\mathcal L}_n}{n!}-\rho\varphi\,,
\end{equation}
where we have defined:
\begin{equation}
    {\mathcal L}_{n+1}=-A_{(2n)}\varphi_{,\mu_1}\varphi_{,\nu_1}\varphi_{,\mu_2\nu_2}...\varphi_{,\mu_n\nu_n}
\end{equation}
\begin{equation}
    A_{(2n)}=-\frac{1}{(5-n)!}\epsilon^{\mu_1,...,\mu_n,\alpha_{n+1},...,\alpha_5}\epsilon^{\nu_1,...\nu_n}_{\alpha_{n+1},...,\alpha_5}\,.
\end{equation}
Note that the $2n$ indices ${\mu_1,\nu_1,...,\mu_n,\nu_n}$ of the tensor $A_{(2n)}$ are not explicitly written for convenience. The variation of the Lagrangians ${\mathcal L}_{n+1}$ gives field equation ${\mathcal E}_{n+1}=0$, where:
\begin{equation}
    {\mathcal E}_{n+1}=(n+1)A_{(2n)}\varphi_{,\mu_1\nu_1}...\varphi_{,\mu_n\nu_n}\,.
\end{equation}
Using these definitions, a simple way to derive that the Lagrangian \eqref{eq:L_MAG} corresponds to the field equation \eqref{eq:MAG_relat} is to develop the determinant in this field equation:
\begin{equation}
\begin{split}
    \text{det}(\eta_{\mu\nu}+\partial_\mu\partial_\nu\varphi)&=\frac{1}{4!}\epsilon^{\mu_1...\mu_4}\epsilon^{\nu_1...\nu_4}\prod_{i=1}^4(\eta_{\mu_i\nu_i}+\partial_{\mu_i}\partial_{\nu_i}\varphi)\\
    &=\frac{1}{4!}\sum_{n=1}^5C^{n-1}_4\frac{(5-n)!}{n}{\mathcal E}_n\\
    &=\sum_{n=1}^5\frac{1}{n!}{\mathcal E}_n\,,
\end{split}
\end{equation}
which shows that the elegant Lagrangian  \eqref{eq:L_MAG} yields the correct equation of motion. The above procedure can be generalized to a curved background.
It is already well-known how to write the Lagrangian in curved space-time for Galileons \cite{gillescedricdeser} which would generate extra interactions with the curvature, however we shall not go further here as the effect is probably negligible for the numerical studies of MAG \cite{MAGus} that we are aiming at explaining in this work.

%% file: 07_Conclusion.tex
In this work we have studied the Monge-Ampère gravity (MAG) as an alternative model of gravity. MAG was proposed by Brenier \cite{brenier_modified_2011} as en effective theory of gravity at cosmological scales. In the Monge-Ampère equation used by Brenier, the potential field is expanded around a harmonic potential \eqref{eq:MA_varphi}. This form of MAG leads to a nonlinear equation relating the determinant of an identity map plus a Hessian to a density field. The main motivation to write MAG in this form is to recover the Newtonian gravity in the weak field limit. However, we have shown that to achieve this task an unusually large background density is needed. If we take such a huge background density then we recover Poisson equation in the weak field limit. However, this huge density is unphysical, being larger that the density of the sun and the planets in the solar system. Furthermore, in this regime, the Optimal Transport formulation of the MA equation at the cosmological scales does not hold. The cosmological interest of MA equation and its link to the optimal transport theory imposes this density to be equal to the background density of the Universe.

We then considered this option. When the density is set to the background density, the Newtonian gravity is recovered in weak field (far field) limit. In the strong gravity regime (near field) the Monge-Ampère gravity is screened and is a constant. The nonlinear terms in the MA equation screen the potential in high-density regions and hence at the solar system scale. This demonstrates that MAG cannot be considered as a replacement for Newtonian gravity but as a model of a new scalar field with Vainshtein-like screening, which arises in theories of modified gravity such as Galileons. The MAG equation then corresponds to a field equation of a scalar field that is coupled to matter, which complements the theory of general relativity.

The original 3-dimensional MA equation \eqref{eq:MA_naive} has relativistic 4-dimensional equivalent, that arises in theories of Galileons and corresponds precisely to the static limit of the field equation of the Lagrangian of quartic Galileon. We have also shown that the MAG equation \eqref{eq:MA_varphi} (where the potential is written in terms of a harmonic potential) is the static limit of the field equation of a particular sum of the Lagrangians of all Galileons in Minkowski space, which can be generalised to curved space.

The rich link between modified theories of gravity, with second-order field equations, {\it e.g.}\ Galileons, and the fast-evolving domain of optimal transport theory uncovered here remains a fertile field of research to further explore.

{\bf Acknowledgement}
Special thanks go to Gilles Esposito-Farèse for numerous advices and discussions on Galileons.
We thank James Binney, Luc Blanchet, Bruno Lévy, André Lukas, Mohammad-Hossein Namdar, Farnik Nikakhtar, Felix Otto, Cale Rankin, Jean-Philippe Uzan, and Cliff Will for discussions.

%% file: 08_Appendix.tex
The second law of Newton for a particle of mass $m$ orbiting around a mass $M$ with a gravitational potential following Monge-Ampère equation gives:
\begin{equation}
    m\frac{\text{d}^2 {\bf{r}}}{\text{d}t^2} = -m(\partial_r\varphi_{\rm N}+\partial_r\varphi_{\rm MA})\hat{\mathbf{r}}\,.
\end{equation}
Since the force is centrifugal, we have the conservation of the angular momentum $h=r^2\dot\theta$, and we can write:
\begin{equation}\label{cent-force}
    \frac{\text{d}^2 {\bf{r}}}{\text{d}t^2} = (\ddot{r}-r\dot{\theta}^{2})\hat{\mathbf{r}} = \left(\ddot{r}-\frac{h^{2}}{r^{3}}\right) \hat{\mathbf{r}}\,.
\end{equation}
With the change of variable $u=1/r$, we can write the second law of Newton as:
\begin{equation}
    \frac{\text{d}^{2} u}{\text{d} \theta^{2}}+u=+\frac{\partial_r\varphi_{\rm N}+\partial_r\varphi_{\rm MA}}{h^{2} u^{2}}\,.
\end{equation}
And using the equation (\ref{eq:MA_solarsystem}), we obtain:
\begin{equation}
    \frac{\text{d}^{2} u}{\text{d}\theta^{2}}+u=\frac{G_{\rm N} M}{h^{2}}+\frac{4\pi G_{\rm MA}\rhobar R}{h^{2}u^2}\,,
\end{equation}
where once again, $R^3=\frac{3M}{4\pi\rhobar}$. We can resolve this equation using perturbation theory. We define the {\it small} parameter $\varepsilon$:
\begin{equation}
    \varepsilon = 4\pi G_{\rm MA}\rhobar R \,,
\end{equation}
\begin{equation}
    \frac{\text{d}^{2} u}{\text{d}\theta^{2}} + u = \frac{G_{\rm N} M}{h^{2}} + \frac{\varepsilon}{h^2u^2}\,,
    \label{eq:NIIepsilon}
\end{equation}
And we look for a solution $u$ such that:
\begin{equation}
    u=u_{0}+\varepsilon u_{1}+\mathcal{O}\left(\varepsilon^{2}\right)\,.
\end{equation}
The zero-th order in $\epsilon$ of the equation (\ref{eq:NIIepsilon}) is:
\begin{equation}
    \frac{\text{d}^{2} u_{0}}{\text{d}\theta^{2}}+u_{0}=\frac{G_{\rm N} M}{h^{2}}\,,
\end{equation}
which is the same as the Newtonian case with the solution:
\begin{equation}
    u_{0}=\frac{G_{\rm N} M}{h^{2}}(1+e \cos \theta)\,.
\end{equation}
The next order term is :
\begin{equation}
    \frac{\text{d}^{2} u_{1}}{\text{d}\theta^{2}} + u_{1}=\frac{1}{h^2u_0^2}\,.
\end{equation}
We can solve this with Mathematica, yielding the following solution :
\begin{equation}
    \begin{split}
         u_1(\theta) = & \frac{h^2}{(G_{\rm N}M)^2}\sin\theta\int_0^\theta\frac{\cos\kappa}{(1+e\cos\kappa)^2}\text{d}\kappa\\
         & -\frac{h^2}{(G_{\rm N}M)^2}\cos\theta\int_0^\theta\frac{\sin\kappa}{(1+e\cos\kappa)^2}\text{d}\kappa\\
    \end{split}\,.
\end{equation}
With this result, and the requirement that at the perihelion $\text{d}u/\text{d}\theta=0$, knowing that :
\begin{equation}
    \begin{split}
        \frac{\text{d}u}{\text{d}\theta} = & - \frac{G_{\rm N}Me}{h^2} \sin\theta \\
        & + \frac{\varepsilon h^2}{(G_{\rm N}M)^2}\sin\theta\int_0^\theta\frac{\sin\kappa}{(1+e\cos\kappa)^2}\text{d}\kappa \\
        & + \frac{\varepsilon h^2}{(G_{\rm N}M)^2}\cos\theta\int_0^\theta\frac{\cos\kappa}{(1+e\cos\kappa)^2}\text{d}\kappa
    \end{split}\,,
\end{equation}
we see that $\theta=0$ is clearly a solution but $\theta=2\pi$ is not and hence we find the answer by looking for the small angle $\delta_{MA}\ll1$ such that $\theta=2\pi+\delta_{MA}$ is a solution :
\begin{equation}
    \delta_{MA}
    =\varepsilon\frac{h^4A(e)}{(G_{\rm N}M)^3e}
    +\mathcal{O}\left(\varepsilon^2\right)\,,
\end{equation}
with $A(e)=\int_0^{2\pi}\frac{\cos\kappa}{(1+e\cos\kappa)^2}\text{d}\kappa=-2\pi e/(1-e^2)^{3/2}$. Using the expression for the angular momentum $h^2=GMr_-(1+e)$, with $r_-$ the perihelion radius, we finally obtain :
\begin{equation}
\delta_{MA}=-8\pi^2\left(\frac{3}{4\pi}\frac{\rhobar^2}{M^2}\right)^{1/3}\frac{G_{\rm MA}}{G_{\rm N}}\frac{r_-^2(1+e)^{1/2}}{(1-e)^{3/2}}\,. 
\end{equation}

%% file: 09_Appendix2.tex
\subsection{Optimal mass transportation and the Monge-Ampère equation: the two legacies of Gaspard Monge}
\label{appendix:MA-OT}

\begin{figure}        
    \includegraphics[scale=0.75]{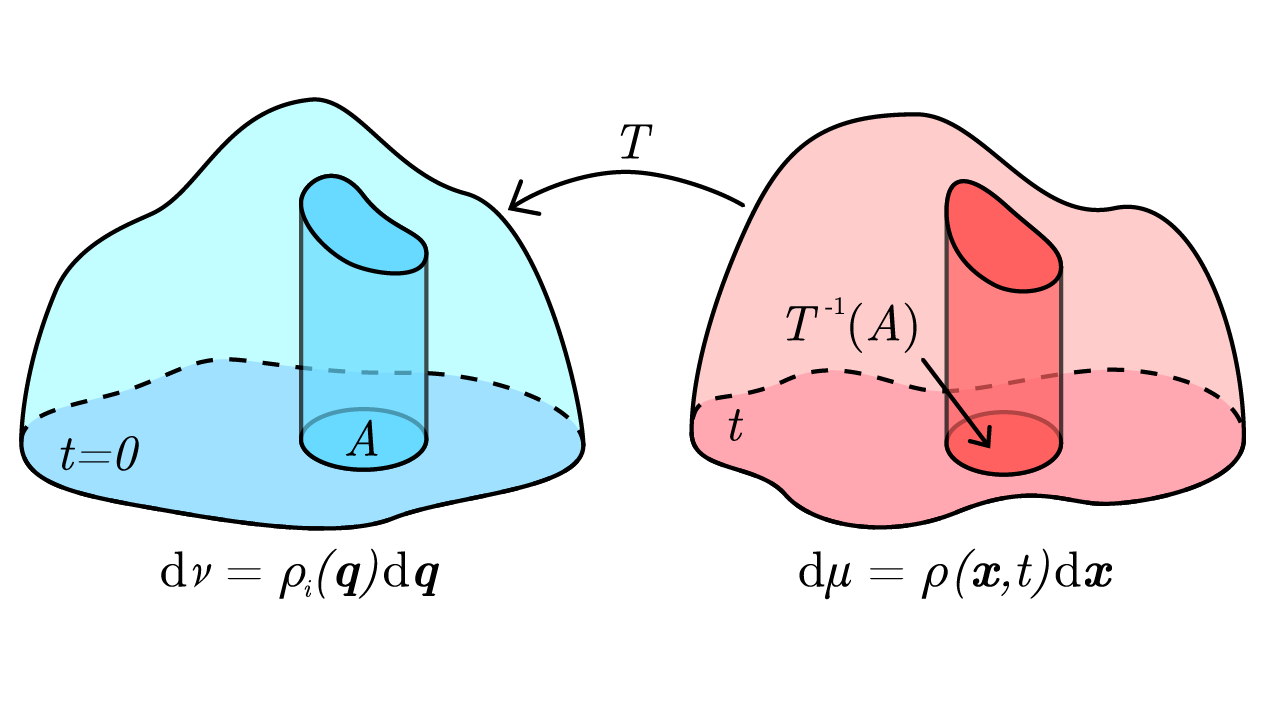}
    \caption{Optimal mass transportation. The optimal transport map $T$ pushes forward the measure $\mu$ into measure $\nu$: $T\#\mu=\nu$, on minimising the cost of transport, C(x,q) which is frequently the square of the Euclidean distance, $\vert \mathbf{x}-\mathbf{q}\vert^2$. The optimal quadratic cost is also known as the Wasserstein or Earth-mover's distance (Figure inspired from \cite{kolouri}).}
    \label{fig:OT-schematic}
\end{figure}

In this appendix, we first show how Monge-Ampère equation appears in cosmology. We then show how it arises mathematically and how it relates to the optimal transport theory.

We wish to describe the evolution of mass density
 in the Universe. We consider a fluid, of initial constant density $\rho_i({\bf q})$ —which is justified as at the beginning of matter-dominated era, apart from tiny fluctuations of the order of $10^{-5}$, the mass density in the Universe was rather uniform— and density field $\rho(\mathbf{x},t)$ at a second chosen instance in time $t$.  The mass conservation equation for a fluid element can be written as:
\begin{equation}
    \rho_i({\bf q})\text{d}\mathbf{q}=\rho(\mathbf{x},t)\text{d}\mathbf{x}\,.
    \label{eq:massconservation}
\end{equation}
With the change of variable $\mathbf{q} \longrightarrow \mathbf{x}$, the volume element becomes:
\begin{equation}
    \text{d}\mathbf{q}=|\text{det}\frac{\partial {\bf q}({\bf x})}{\partial {\bf x}}|\text{d}\mathbf{x}\,.
    \label{eq:pre-MA-cos}
\end{equation}
 Supposing that the fluid is curl-free-- a valid assumption in an expanding Universe where vorticity modes decay--then we can write  the mapping between initial and final positions as the gradient of a potential $\varepsilon$, {\it i.e.} ${\bf x}({\bf q})=\nabla_{\bf q}\varepsilon({\bf q})$. Then by Legendre-Fenchel transform we can also write the inverse map
 $\mathbf{q}(\mathbf{x})=\nabla_{\bf x}\Phi(\mathbf{x})$, where both forward, $\varepsilon$, and backward potentials, $\Phi$, are convex. Substituting for the map in \eqref{eq:pre-MA-cos} we obtain the Monge-Ampère equation:
\begin{equation}
    \text{det}(D^2\Phi)=\rho(\mathbf{x},t)/\rho_i({\bf q})
    \label{eq:MA}\,.
\end{equation}

In astrophysics, the approximation $\bf{x}(\mathbf{q})=\nabla\varepsilon(\mathbf{q})$ can be shown to be valid for describing the Universe on very large scales where vorticity modes vanish and multistream regions in space are insignificant (see \cite{brenier_reconstruction_2003} and references therein). If the potential $\Phi$ is expressed in terms of a harmonic potential, {\it i.e.} $\Phi={\bf q}^2/2+\varphi$, we obtain the equation of Monge-Ampère gravity (MAG) \eqref{eq:MA_varphi},  \cite{brenier_modified_2011,brenier_double_2015}, which we write here again as:
\begin{equation}
    \text{det}(\mathbb{I}+D^2\varphi)=\rho(\mathbf{x},t)/\rho_i\,.
    \label{eq:MA-fast}
\end{equation}

Therefore, this procedure shows that when one makes parallels between the Poisson and the Monge-Ampère equation, one is essentially comparing the Jacobian equation for a gradient velocity field ({\it i.e.} the Monge-Ampère equation) to the Poisson equation.

The Monge-Ampère equation is nonlinear and difficult to solve. In this paper we have analytically solved the Monge-Ampère equation for the special case of spherically symmetric systems. An elegant method for solving MA equation is provided via optimal transport theory (see {\it e.g.} \cite{peyre-cuturi,journals/M2AN/LevyNAL15,levy_mnras_2021}) which we shall briefly outline below. 

We have so far shown how the Monge-Ampère equation arises in astrophysics as a consequence of mass conservation and because the Universe is modeled as a fluid whose displacement field can be well-approximated by the gradient of a convex potential. Below we make a formal mathematical derivation of Monge-Ampère equation and show how it naturally links to the optimal mass transportation problem.

The optimal mass transport theory seeks to transform one probability distribution to another, under the condition that a given cost function, $C({\bf q},{\bf x})$, is minimized. The problem was initially studied by the French Mathematician Gaspard Monge and presented in his seminal work “Mémoire sur la théorie des déblais et des remblais” \cite{monge} in 1781. It was for military reasons, namely the construction of fortifications, that Monge wanted to know how to move, in the Euclidean space , the soil (déblais) to piles or fortification sites (remblais) while minimizing the {\it cost} of transportation. A historical account of the work of Monge in this context is presented in the next section of this Appendix.

 A schematic presentation of Monge's problem is given in Fig.\ \ref{fig:OT-schematic} and its explanation is as follows. 
Given two probability measures, {\it e.g.} in Fig.\ \ref{fig:OT-schematic}, these are $\rho_i({\bf q}) {\rm d}{\bf q}$ and $\rho({\bf x}) {\rm d}{\bf x}$, one wishes to morph one distribution to the other by optimizing a {\it cost function}. Note that  $\rho_i({\bf q})$ and $\rho({\bf x})$ can be at two different time slices, which is often the case in astrophysics. Hereafter, we drop the time variable for the densities and write $\rho_i({\bf q})$ and $\rho({\bf x})$ for the two density fields.

In the original formulation of Monge, the cost to transport molecules of mass $m$ —as he refers to it in his paper \cite{monge}— from ${\bf q}$ to ${\bf x}$ was the mass multiplied by the Euclidean distance; $m\,\vert {\bf x} - {\bf q}\vert $. The transport was also constrained to conserve mass, {\it i.e.} $\rho_i({\bf q}) {\rm d}{\bf q}=\rho({\bf x}) {\rm d}{\bf x}$. 
 
One need not be restricted to Euclidean distance and more general cost functions $C({\bf x}, {\bf q})$ can be used. Indeed Monge's cost function is ill-defined, as there are cases where the problem is without a solution or with multiple solutions. There is also a long list of efforts made to prove the existence of an optimal map for Monge's cost function (see {\it e.g.}\cite{sudakov,gangbo,cfm}).
 
In the modern formulation of Monge's problem with a general cost function one searches for a map, $T({\bf x})$ such that one probability measure, ${\rm d}\mu({\bf x})=\rho({\bf x}) {\rm d}{\bf x}$ is {\it pushed forward} into another probability measure, {\it e.g.} ${\rm d}\nu({\bf q})=\rho({\bf q}) {\rm d}{\bf q}$. Using the notations for push forward, $\#$, we can write the morphing of one probability distribution into another one as
\begin{equation}
    T\#\mu=\nu\qquad\qquad \mu({\bf x})=\nu(T^{-1}({\bf x}))\,,
\end{equation}
and this map $T({\bf x})$, is the optimal transport map. That is for all different maps (optimal and non-optimal) $M\#\mu=\nu$, it is the map $T$ that minimizes the transport cost. The optimal map $T$ has to be a diffeomorphism, {\it i.e.}  $T$ and $T^{-1}$ should be differentiable.

Let us call this optimal action $I$ on which the minimization is performed
\begin{equation}
 I=\underset{\mathbb{R}^3}{\int}  C(T({\bf x}),{\bf x}) {\rm d}\mu({\bf x})=\underset{M\# \mu=\nu}{\rm min}\underset{\mathbb{R}^3}{\int} C(M({\bf x}),{\bf x}) {\rm d}\mu({\bf x})\,.
   \label{eq:cost-OT}
\end{equation}
The mapping has to be measure-preserving, that is it has to conserve mass,
\begin{equation}
\underset{\mathbb{R}^3}{\int} \rho_i({\bf q}) {\rm d}{\bf q}=\underset{\mathbb{R}^3}{\int} \rho({\bf x})\, {\rm d}{\bf x}\,.
\label{eq:mass-conservation}
\end{equation}
It is a normal practice to add the constraint equation \eqref{eq:cost-OT} to the action through a Lagrange multiplier. Therefore, the constrained optimal action $I_c$ is
\begin{eqnarray}
I_c&=&\underset{M\# \mu=\nu,\Theta}{\rm min}\left[\underset{\mathbb{R}^3}{\int} C(M({\bf x}),{\bf x}) {\rm d}\mu({\bf x})\nonumber\right .\\
&+&\left .\underset{\mathbb{R}^3}{\int} \Theta({\bf q})\,\rho_i({\bf q}) {\rm d}{\bf q}-\Theta(M({\bf x}))\rho({\bf x}))\, {\rm d}{\bf x}\right]\,.
   \label{eq:constrained-action}   
\end{eqnarray}

As we have mentioned the Monge's initial cost function, $C(M({\bf x}),{\bf x})$ used a Euclidean distance $\vert {\bf x}-{\bf q}\vert$ which is ill-defined. The solution to the Monge's mass transportation, which dates back to 200 years ago was recently found by Yann Brenier \cite{yann_decompositionpolair,yann_decompositionpolair2}, who proposed that for a quadratic cost function, $\vert {\bf x}-{\bf q}\vert^2$, the optimal map $T$ is unique and is given by the gradient of a convex function, {\it i.e.}, $T=\nabla\Theta$. We shall provide a sketch of the proof here. (For full mathematical proof, the reader is referred to the original papers by Brenier \cite{yann_decompositionpolair,yann_decompositionpolair2}.) The optimal quadratic cost, corresponding to the optimal map $T$, is also well-known as the Wasserstein or Earth-mover's distance.

For a quadratic cost function, the expression \eqref{eq:constrained-action} reads
\begin{eqnarray}
I_c&=&\underset{M\# \mu=\nu,\Theta}{\rm min}\left[\underset{\mathbb{R}^3}{\int} \frac{\vert {\bf x}-M({\bf x})\vert^2}{2}  {\rm d}\mu({\bf x})\nonumber\right .\\
&+&\left .\underset{\mathbb{R}^3}{\int} \Theta({\bf q})\,\rho_i({\bf q}) {\rm d}{\bf q}-\Theta(M({\bf x}))\rho({\bf x}))\, {\rm d}{\bf x}\right] \,.
   \label{eq:constrained-action-quadratic}   
\end{eqnarray}
It is important to note that 
\begin{eqnarray}
  \underset{M\# \mu=\nu}{\rm min}\underset{\mathbb{R}^3}{\int}
   \frac{\vert {\bf x}-M({\bf x})\vert^2}{2} {\rm d}\mu({\bf x})  &=& \underset{M\# \mu=\nu}{\rm min}\underset{\mathbb{R}^3}{\int}(-{\bf x}\cdot M({{\bf x}}))\, {\rm d}\mu
   \nonumber\\
   &=&\underset{M\# \mu=\nu}{\rm sup}\underset{\mathbb{R}^3}{\int}({\bf x}\cdot M({{\bf x}}))\, {\rm d}\mu \nonumber\\
\end{eqnarray}
Hence, the expression for constrained action with a quadratic cost function, \eqref{eq:constrained-action-quadratic} becomes
 \begin{eqnarray}
I_c&=&\underset{M\# \mu=\nu}{\rm sup}\underset{\Theta}{\rm min}\left[\underset{\mathbb{R}^3}{\int} { {\bf x}\cdot M({\bf x})}\,  {\rm d}\mu({\bf x})\nonumber\right .\\
&+&\left .\underset{\mathbb{R}^3}{\int}\Theta({\bf q})\,\rho_i({\bf q}) {\rm d}{\bf q}-\Theta(M({\bf x}))\rho({\bf x}))\, {\rm d}{\bf x}\right] \,.
   \label{eq:constrained-action-quadratic-2}   
\end{eqnarray}
Next we apply the variational principle on the Lagrangian density, integrand of the above action, recalling that ${\rm d}\mu=\rho({\bf x}) {\rm d}{\bf x}$, 
\begin{eqnarray}
 \frac{\partial {\cal L}}{\partial M}&=&{\bf x}\rho({\bf x}) -{\nabla_M}\Theta(M({\bf x}))\rho({\bf x}) \nonumber\\
 &=& 0\nonumber\\
 &\Rightarrow &{\bf x}= \nabla\Theta(T({\bf x}))\,,
 \label{eq:optimality1}
\end{eqnarray}
which means that we have a gradient mapping. Note that as the above gives the optimal mapping, in the last line we have replaced $M$ by $T$. We emphasise that in the last line the $\nabla$ is taken with respect to $T({\bf x})$ which is our Lagrangian coordinate ${\bf q}$.

It is important to note that the optimal map is monotone (crossing of "trajectories" or taking-over in one  dimension, is not allowed). The monotonicity condition can be written as

\begin{equation}
({\bf x_2}-{\bf x_1})\cdot (T({\bf x}_2)-T({\bf x_1})) > 0\,,
\end{equation}
which is replaced by the stronger condition of {\it cyclic monotonicity} in more than one dimension. Cyclic monotonicity is tightly related to the convexity and the uniqueness of the optimal map. Indeed, for the optimal solution given above, the second variation of the Lagrangian density will inform us about the nature of the potential $\Theta$. That is
\begin{eqnarray}   
\frac{\partial^2 {\cal L}}{\partial M^2}&=&\nabla^2\Theta\\
&=& 0\nonumber\\
   &\Rightarrow & \Theta \quad \text{is a convex function.}
   \label{eq:optimality2}
\end{eqnarray}

Having derived the optimality conditions, namely \eqref{eq:optimality1} and \eqref{eq:optimality2}, we can go back to our equation of mass conservation \eqref{eq:mass-conservation} which can be written for any test function, $\xi$ as
\begin{equation}
\underset{\mathbb{R}^3}{\int}  \xi({\bf q}) \rho_i({\bf q}) {\rm d}{\bf q}=\underset{\mathbb{R}^3}{\int} \xi(M({\bf x}))\rho({\bf x})\, {\rm d}{\bf x} \,.
\end{equation}
As before, we use the definition ${\bf q} = M({\bf x})$ and hence
\begin{equation}
\underset{\mathbb{R}^3}{\int}\xi(M({\bf x})) \rho_i(M({\bf x})) \vert {\rm det}\nabla M({\bf x})\vert {\rm d}{\bf x} =\underset{\mathbb{R}^3}{\int}\xi(M({\bf x}))\rho({\bf x}) {\rm d}{\bf x}
\end{equation}
Note that we are working in $\mathbb{R}^3$ but the results here are valid in any dimensions. Hence we obtain 
\begin{equation}
\vert {\rm det}\nabla M({\bf x})\vert=\frac{\rho({\bf x})}{\rho_i(M({\bf x}))}\,,
\label{eq:pre-MA}
\end{equation}
which is a sort of pre-Monge-Ampère equation because we have the gradient of a mapping and not the Hessian of a potential as appears in the Monge-Ampère equation. We can use \eqref{eq:optimality1} to find the inverse map, needed in \eqref{eq:pre-MA}, via a Legendre-Fenchel transform, 
\begin{eqnarray}
    \Theta({\bf x})&=&\underset{\bf q}{\rm sup}\left[{\bf x}\cdot {\bf q}-\vartheta({\bf q}))\right]\\
    \vartheta({\bf q})&=&\underset{\bf x}{\rm sup}\left[{\bf x}\cdot {\bf q}-\Theta({\bf x})\right] 
\end{eqnarray}
Thus if 
\begin{equation}
    {\bf x}({\bf q})=\nabla_{\bf q}\Theta({\bf q})\,,
\end{equation}
then the inverse map is also a gradient of another potential, $\vartheta$, which is also convex and hence we can write
\begin{equation}
    {\bf q}({\bf x})=\nabla_{\bf x}\vartheta({\bf x})\,.
\end{equation}
Hence in expression \eqref{eq:pre-MA} the Hessian clearly emerges using the optimality condition \eqref{eq:optimality1} and the above Legendre-Fenchel transform, to yield
\begin{equation}
    \vert {\rm det} \nabla^2\Theta({\bf x})\vert =\frac{\rho({\bf x})}{\rho_i(T({\bf x}))}\,,
\end{equation}
which is the Monge-Ampère equation as we have obtained earlier in the context of mass-conservation in the Universe \eqref{eq:MA-fast}.

We can solve the Monge-Ampère equation, and find the potential $\Theta$, directly using  existing methods of solving partial-differential equations such as viscosity method or finite-element method \cite {finiteelement,viscosity,2solutions}. These methods are mostly inefficient as MA equation is highly nonlinear. Care also has to be taken to ensure uniqueness of solutions when employing different numerical schemes. The other possibility is, as we have shown, to solve the optimal transport problem, {\it i.e.} to solve the variational problem presented in \eqref{eq:constrained-action} through optimal transport theory. 

Here we outline the progress in solving the MA equation through optimal transport. The constrained action in \eqref{eq:constrained-action-quadratic} can be simply discretized if we consider that a continuous spatial mass distribution can be approximate by  $N$ identical Dirac particles. Thus our initial and final densities, $\rho_i({\bf q})$ and $\rho({\bf x})$ respectively, can be written as
\begin{equation}
    \rho_i({\bf q})=\sum_{j=1}^N\delta({\bf q}- {\bf q}_j)\,,\qquad 
    \rho({\bf x})=\sum_{k=1}^N\delta({\bf x}- {\bf x}_k)\,,
\end{equation}
where we recall that index $i$ in $\rho_i$ stands for initial and is not particle number. As for discrete densities, to verify the mass conservation constraint, the map $T({\bf x})$ has to induce a one-to-one matching between particle positions in the final space (${\bf x}$) and initial space (${\bf q}$), which may be written as a permutation of indices that sends ${\bf x}_k$ to ${\bf q}_j$. Hence the minimization of the quadratic cost function in \eqref{eq:constrained-action} can be written as
\begin{equation}
    I_{\rm optimal}=\underset{j}{\rm min}\sum_k^N \frac{\vert {\bf x}_k- {\bf q}_{j(k)}\vert^2}{2}\,,
\end{equation}
and the mass conservation is satisfied by construction. We are now face at a purely combinatorial problem —an instance of assignment problem— of finding a permutation $j(k)$ that minimizes the above quadratic cost function. 

For N particles, the above assignment problem has a single solution which can be found by going through all of the $ N\!$ permutations. However, it is well-known that the complexity of the  assignment problem can be reduced to {\it polynomial time} through the dual formulation, based on Kantorovich relaxation which then reduces the problem to one in linear programming \cite{brenier_reconstruction_2003, bertsekas}.
Here we shall not go through the details of discrete-discrete algorithms (the reader is referred to {\it e.g.} \cite{bertsekas,peyre-cuturi} but mention the {\it semi-discrete} algorithm which we have developed to tackle the cosmological reconstruction problem \cite{brunoroyasebastian} and which we have also used in simulating Monge-Ampère gravity, as shown in Fig.\  \ref{fig:mag-newton}, and explained in details in our companion paper \cite{MAGus}. We had briefly discussed the idea of a semi-discrete optimal transport many years ago in \cite{brenier_reconstruction_2003}, and mentioned at the time that we could not find an algorithm. The breakthrough came years later when a clear algorithm was set out for semi-discrete optimal transport \cite{merigot,kitagawa,journals/M2AN/LevyNAL15} and we then adapted it to cosmological reconstruction problem \cite{brunoroyasebastian}.

Here, we briefly discuss semi-discrete optimal transport algorithm. So far, we have discretized the mass both at the initial and final times. However, we can consider a different problem in which we only discretize the present density field (${\bf x}$) and leave the initial density field (${\bf q}$) continuous. This is justified in cosmology, as the early Universe is rather homogeneous and can be well-approximated by a uniform density field, whereas galaxies in the late Universe, as tracers of dark matter field, are discrete. We can thus consider {\it a semi-discrete} setting. In semi-discrete optimal transport, finding the optimal assignment is replaced by the tessellation of the initial uniform space into polyhedron. The reason why these regions are polyhedron is that the convex potential of the Lagrangian map has a gradient that takes only finitely many values. This problem was initially studied by Aleksandrov and Pogorelov (see, {\it e.g.} \cite{Pogorelov}), and is related to the famous Minkowski problem in mathematics \cite{minkowski}. The problem is to construct a convex polyhedron with prescribed areas and orientations. However these beautiful mathematical ideas only became concrete in form of algorithms only recently \cite{merigot, journals/M2AN/LevyNAL15}. These polyhedron are referred to as the Laguerre cells, and each cell at initial time is assigned to a galaxy at the present time (or any other later time). The optimal mapping between the two is the unique solution to a semi-discrete
optimization problem \cite{brunoroyasebastian} and is specified by that
set of $\Theta_k$ values which maximizes the {\it Kantorovich dual}
\begin{equation}
    K(\Theta)=\sum_k\int_{V_k^\Theta}
    \left[
    \frac{1}{2}\vert {\bf x}_k-{\bf q}\vert^2-\Theta_k\right] {\rm d}{\bf q}+\sum_k\ v_k\Theta_k\,,
\end{equation}
through Lagrange multiplier, $\Theta$, and $v_k$ is the volume of the $k$-th Laguerre cell, imposed as a constraint (the $v_k$ sum up to the total volume $V$). If all $\Theta_k$ are equal, the Laguerre diagram reduces to Voronoi tessellation.
An efficient and convergent Newton method \cite{kitagawa} is then used to find the optimal potential which makes a huge gain in computational speed (complexity is of the order of $N {\rm log} N$ with $N$ the number of Laguerre cells \cite{brunoroyasebastian}) as compared to initial brute combinatoric problem with complexity of $N!$ and the discrete-discrete algorithms based on Kantorovich duality with polynomial complexity.
For a full account of semi-discrete optimal transport the reader is referred to {\it e.g.} \cite{journals/M2AN/LevyNAL15,brunoroyasebastian} and references therein.

Indeed, there has been various breakthroughs also in the development of discrete-discrete algorithms in the last decade which also explains the recent surge in the applications of optimal transport to diverse areas of science (see {\it e.g.} \cite{peyre-cuturi, brunoroyasebastian,bertsekas,cuturi,journals/M2AN/LevyNAL15}).

\subsection{Historical note on Gaspard Monge, optimal mass transportation and Monge-Ampère equation}
\label{appendix:history}
Gaspard Monge, Comte de Pélouse, was born on the 9th of May 1746 and lived till 28 July 1818. He was a contemporary of Euler, Lagrange, Laplace and Gauss. A French mathematician, engineer and above all a geometer, Monge was not from a family of Nobles but worked in the prestigious École de Mézières, reserved for the aristocrats, because of his genius. The school, among others, was specialised in the construction of fortifications. This is how Monge laid out the famous problem of mass transportation about 10 years before the French revolution. Contrary to his contemporaries like Lagrange, Laplace or Legendre, Monge became actively involved in the French revolution. He became a close friend of Napoleon Bonaparte and accompanies him in various expeditions and especially in  his famous trip to Egypt.  Monge founded the French École Polytechnique in 1794 and was made the count of Pélouse (a city in antique Egypt) by Napoleon Bonaparte in 1808. After the downfall of Napoleon and restoration of monarchy in 1815, Monge was stripped of all his honours and positions and was discharged from the École Polytechnique, and later died in isolation in 1818.

Monge is considered the father of differential geometry and is one of the discoverers of partial differential equations ({\it e.g.}\ see \cite{Monge-hist}). Here,  we shall only give a short historical account of his mass transportation problem and his nonlinear partial differential equation, the Monge-Ampère equation.

Monge presented his mass transportation problem in 1776 in the academy of science and under the encouragement of Condorcet, permanent secretary of the academy of science, published his famous article, entitled {\it Théorie des déblais et des remblais} a few years later in 1781. This remarkable article is considered a starting point for optimal transport theory and also the corner stone of differential geometry. 

In his article, motivated by the construction of fortifications, Monge laid out the optimal mass transportation problem : {\it how can one transfer "molecules of soil" from déblais (piles) to remblais (holes) by minimizing the cost of transfer ?}. He wrote the cost of transfer as the Euclidean distance multiplied by the mass \cite{monge,kantormonge,villani1,villani2}. In his paper, Monge does not provide a solution to the mass transportation problem, however he provides important geometrical hints. Monge conjectures that when the cost of transport is the Euclidean distance then there exists an optimal map that transports the pile of soil to holes. He states that a property of the optimal map is that the transport routes do not cross. In the modern language of fluid mechanics and astrophysics one says that there is no shell-crossing or multi-streaming and in the language of mathematics, one says that the map is convex. In modern terms, the map is said to be {\it monotone} in one dimension and in higher dimensions one often uses the term {\it cyclic monotonicity} to impose the absence of crossings. 

In his article, Monge affirms that the best transport map from one domain of space in 3 dimensions towards another domain in 3 dimensions is traced along the family of normal to a "certain surface". The quest was then to find this surface and the search for this surface lies at the foundation of his nonlinear partial differential equation, in 1784 (without a Hessian), which was later generalised by Ampère (with a Hessian) in 1820, and which is now known as the Monge-Ampère equation.

Thus in spite of its apparent simplicity, Monge's problem remained unsolved for centuries. The problem was considered so important that a century later the academy of science put a Bordin prize on it in 1884, of a value of 3000 French Francs. Paul Appell, a well-known French mathematician and a contemporary of Henri Poincaré, finds the link between Monge’s problem and certain surfaces with geometrical properties, such as minimal surfaces which are nowadays called Appell surfaces. The {\it mémoire} written by Appell then won the prestigious Bordin prize \cite{mongeAppell,gheys1,gheys2}.

However Appell did not really solve the mass transportation problem and it was not until two hundred years later that Leonid Kantorovich, a Russian mathematician and economist, unaware of Monge's work, proposed an algorithm for finding not an optimal transport map but an optimal transport plan in 1942 \cite{kantor42}. About six years later, Kantorovich was made aware of Monge’s mass transportation problem and made the connection between his work and that of Gaspard Monge in a short paper entitled; {\it A Problem of Monge} \cite{kantormonge}.

In Monge's formulation, there is a one-to-one correspondence between source and target measures, {\it i.e. in Monge's formulation} between holes and piles. In Kantorovich's formulation, a pile can split and go to different holes and conversely a single hole can receive soil from different piles and hence one works with product space of "déblai-remblai". Not only Kantorovich {\it relaxed} Monge's problem, allowed for multi-streaming,  and represented a probabilistic version of Monge's mass transportation problem but he also laid the foundation for numerical solutions of the optimal mass transportation by evoking {\it duality}. Kantorovich’s problem, is simpler than Monge’s one. In Kantorovich transportation problem one searches for a measure  say $\gamma$ in the product space ${\bf q}\times {\bf x}$ ({\it i.e.} in our example $\gamma$ has $\mu$ and $\nu$ as marginals) and minimises the transport cost $C_k(\gamma)=\int C(x,q)d\gamma(x,q)$ for all choices of $\gamma$. In Kantorovich's formulation one solves a {\it linear programming} and this is why Kantorovich is often referred to as the father of  linear programming. In the light of the seminal works by Kantorovich, Monge's mass transportation problem is now-a-days known as Monge-Kantorovich problem.

However, it is only about 30 years ago that the relation between Monge's mass transportation and the Monge–Ampère equation was unravelled \cite{yann_decompositionpolair,yann_decompositionpolair2}. Yann Brenier showed the equivalence of the elliptic Monge–Ampère equation and Monge's mass transportation problem with quadratic cost: when initial and final distributions are well-defined, the optimal solution is actually one-to-one, in spite of Kantorovich relaxation \cite{yann_decompositionpolair,yann_decompositionpolair2} and the optimal map is given by the gradient of a convex potential (Kantorovich potential) that also solves the Monge-Ampère equation (for mathematical details see the previous section in this Appendix). The result of Brenier was later extended to general cost functions \cite{gangbomccann}.

It is interesting that Monge, Lagrange and Euler were contemporaries. Indeed Monge and Lagrange were both members of the prestigious {\it Comité des poids et mesures} which invented standard units: meter and grams among others. Monge's mass transportation problem is similar to another optimisation problem frequently encountered in physics, namely the Euler-Lagrange's action optimisation principle. 

Is there a relation between Monge's mass transportation problem and the Euler-Lagrange action minimisation ? The former considers an optimal matching between different probability distributions and the latter considers minimization of trajectories between fixed boundaries (with known matching). One can consider the latter as a dynamical problem and the former as a static problem (there is no time in Monge's mass transportation problem).

The link between the Monge's mass-transportation and variational principle was formulated mathematically only recently in 2000 \cite{benamoubrenier}. In this seminal paper, it was shown that the optimisation with a quadratic cost, written in Eulerian coordinates, is similar to the minimisation of Euler-Lagrange action with only a kinetic term.